\documentclass[preprint,12pt,sort&compress]{elsarticle}

\usepackage{amssymb}
\usepackage{natbib}
\usepackage{graphicx}
\usepackage{color} 
\usepackage{hyperref}
\usepackage{epstopdf, epsfig}
\usepackage{caption}
\usepackage{subcaption}
\usepackage[utf8]{inputenc}

\journal{Fusion Engineering and Design}

\begin{document}

\begin{frontmatter}



\title{Feasibility of the EDICAM camera for runaway electron detection in JT-60SA disruptions}


\author[label1,label2]{Soma Olasz}
\ead{olaszsoma@reak.bme.hu}
\author[label3]{Mathias Hoppe}
\author[label2]{Tamás Szepesi}
\author[label4]{Kensaku Kamiya}
\author[label1,label2]{Peter Balazs}
\author[label1,label2]{Gergo I. Pokol}

\affiliation[label1]{organization={Institute of Nuclear Techniques, Faculty of Natural Sciences, Budapest University of Technology and Economics},
               addressline={Muegyetem rakpart 3.},
               city={Budapest},
               postcode={1111},
               country={Hungary}}

\affiliation[label2]{organization={Fusion Plasma Physics Department, Centre For Energy Research},
               addressline={Konkoly-Thege Miklós út 29-33},
               city={Budapest},
               postcode={1121},
               country={Hungary}}

\affiliation[label3]{organization={Ecole Polytechnique Fédérale de Lausanne, Swiss Plasma Center},
               city={Lausanne},
               postcode={CH-1015},
               country={Switzerland}}

\affiliation[label4]{organization={National Institutes for Quantum and Radiological Science and Technology},
               city={Naka},
               postcode={311-0193},
               country={Japan}}

\begin{abstract}
The visible camera system EDICAM (Event Detection Intelligent Camera), recently installed on JT-60SA, is simulated to assess whether it can be used for measuring synchrotron radiation from relativistic runaway electrons. In this simulation, the SOFT synthetic synchrotron diagnostic framework is used to compute the synthetic synchrotron camera images from a JT-60SA-like disruption modelled with the DREAM disruption simulation code. In the studied scenario, a large amount of argon is added to the plasma, and a disruption is simulated by starting a prescribed exponential temperature drop and finishing with further cooling provided by the argon in a self-consistent simulation of the current quench. The background plasma evolution is calculated by DREAM self-consistently with the fast electron population, which is modelled kinetically. The resulting runaway electron distribution function along with the parameters of the EDICAM visible camera system are used as an input to the SOFT framework to assess the feasibility of the camera for runaway electron detection. We find that the runaway electron beam formed in the disruption can produce synchrotron radiation observable by the EDICAM system, thus enabling the use of the EDICAM for the characterization of runaway electron beams.
\end{abstract}


\end{frontmatter}


\section{Introduction}\label{sec:intro}
Runaway electron generation has been extensively studied in recent years~\cite{Paz_Soldan21, Reux21, Pautasso20, Hollmann15, Zeng13} due to the major threat they pose for future large-scale experiments~\cite{Boozer17, Lehnen15}. The generated runaway current is exponentially dependent on the pre-disruption plasma current due to the avalanching effect~\cite{Rosenbluth97, Smith06}, so in future large current devices, the problem of runaway electrons is expected to be more severe than in present day tokamaks. Extensive study of runaway electrons is hence important.
 
Runaway electrons can emit various forms of radiation, and this is used for their detection in current devices~\cite{Pace16, Cerovsky22}. Bremsstrahlung radiation is produced due to the interaction of runaway electrons and other plasma particles and is typically measured up to the order of $\rm MeV$ photon energy~\cite{Pace16}. It is routinely used for the detection of runaway electron presence in a tokamak. Synchrotron radiation can be detected by visible and infrared camera systems~\cite{Wongrach14, Popovic21} and it can be used to gather information on the electron distribution function -- both in real space and momentum space~\cite{Tinguely18, Hoppe18, Wijkamp21, Carbajal17}.

Synchrotron radiation images on various devices and modelling of runaway electron synchrotron radiation were used in synthetic diagnostic tools to get information on the runaway electron distribution function. One such study was done with the KORC code~\cite{Carbajal17}, which has been used to study the relation between the characteristics of the runaway electron population and the background plasma parameters. These simulations include full orbit effects of the electrons. The SOFT synthetic diagnostic framework~\cite{Hoppe18_2}, which works in guiding centre coordinates, has been used to model synchrotron radiation on several tokamaks, including Alcator C-Mod, DIII-D and TCV.~\cite{Tinguely18, Hoppe18, Wijkamp21}.

The JT-60SA tokamak is planned to begin operation soon and it is expected to contribute to runaway electron mitigation studies for ITER and DEMO~\cite{Giruzzi19}. In this paper, we assess the feasibility of the recently installed Event Detection Intelligent Camera (EDICAM) visible camera system on JT-60SA~\cite{Szepesi20} for runaway electron detection and characterization. The EDICAM system has not been considered for runaway electron detection before, and we propose the use of this visible camera for this purpose as no infra-red cameras are dedicated for runaway electron detection on JT-60SA. The EDICAM parameters were given to the SOFT framework, which was used to simulate a synthetic synchrotron radiation image from a runaway electron distribution generated in a JT-60SA-like disruption. The disruption was simulated using the DREAM disruption runaway electron modelling tool~\cite{Hoppe21}.

The characteristics of the EDICAM system are introduced in section~\ref{sec:edicam}. A description of the modelling tools used in the simulations is given in section~\ref{sec:models}, while the simulation settings, results, and their interpretation are presented in section~\ref{sec:Results}. Finally, we conclude in section~\ref{sec:Conclusions}.

\section{The EDICAM system}\label{sec:edicam}
The parameters of the EDICAM diagnostic are discussed in this section, with a focus on the camera characteristics, such as location, field of view, and sensitivity range, required for the simulation.

The JT-60SA tokamak is divided to 18 different sectors, and an EDICAM visible camera system is installed in the P18 sector of the device~\cite{Giruzzi19}. It features non-destructive read-out, meaning it can read out data from the sensors without affecting the exposure process, which enables simultaneous observation of multiple different regions in the camera field of view, specific regions-of-interest (ROI), as well as the full field of view (FOV), with different sampling rates. This makes it ideal for the observation of various fast phenomena, including some phases of disruptions~\cite{Szepesi20}.

\begin{figure*}
\begin{center}
\includegraphics[scale=0.3]{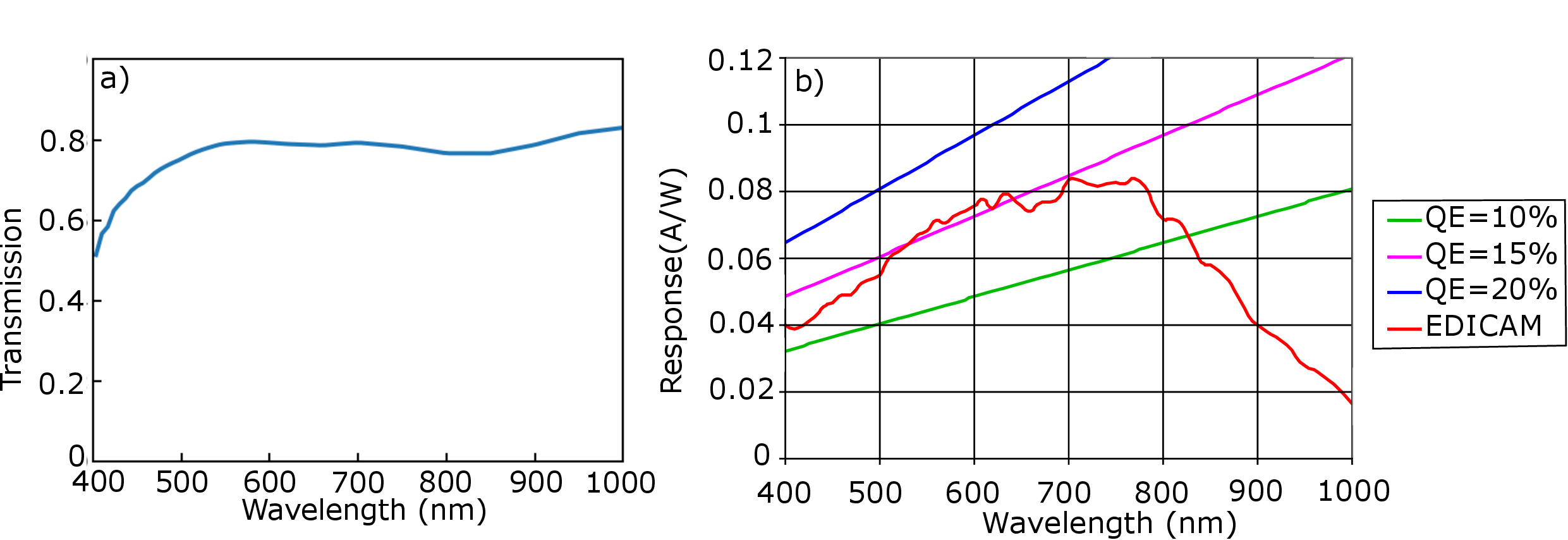}
\caption{The spectral sensitivity of the EDICAM camera system. Fig. 1(a) shows the transmission efficiency of the optical elements as a function of wavelength. The optical elements transmit light with high efficiency above $~500\; \rm nm$. Fig. 2(b) shows the response of the camera sensor as a function of wavelength. The quantum efficiency (QE) is the fraction of the photon energy absorbed by the sensors including the effect of non-sensitive areas of the pixels. The lines shown correspond to $10$, $15$ and $20 \%$ efficiency. The EDICAM system sensors have a quantum efficiency of about $15 \%$ between $500$ and $700\; \rm nm$, but it can detect photons up to $1000\; \rm nm$.}
\label{spectral}
\end{center}
\end{figure*}

\begin{figure*}
\begin{center}
\includegraphics[scale=0.3]{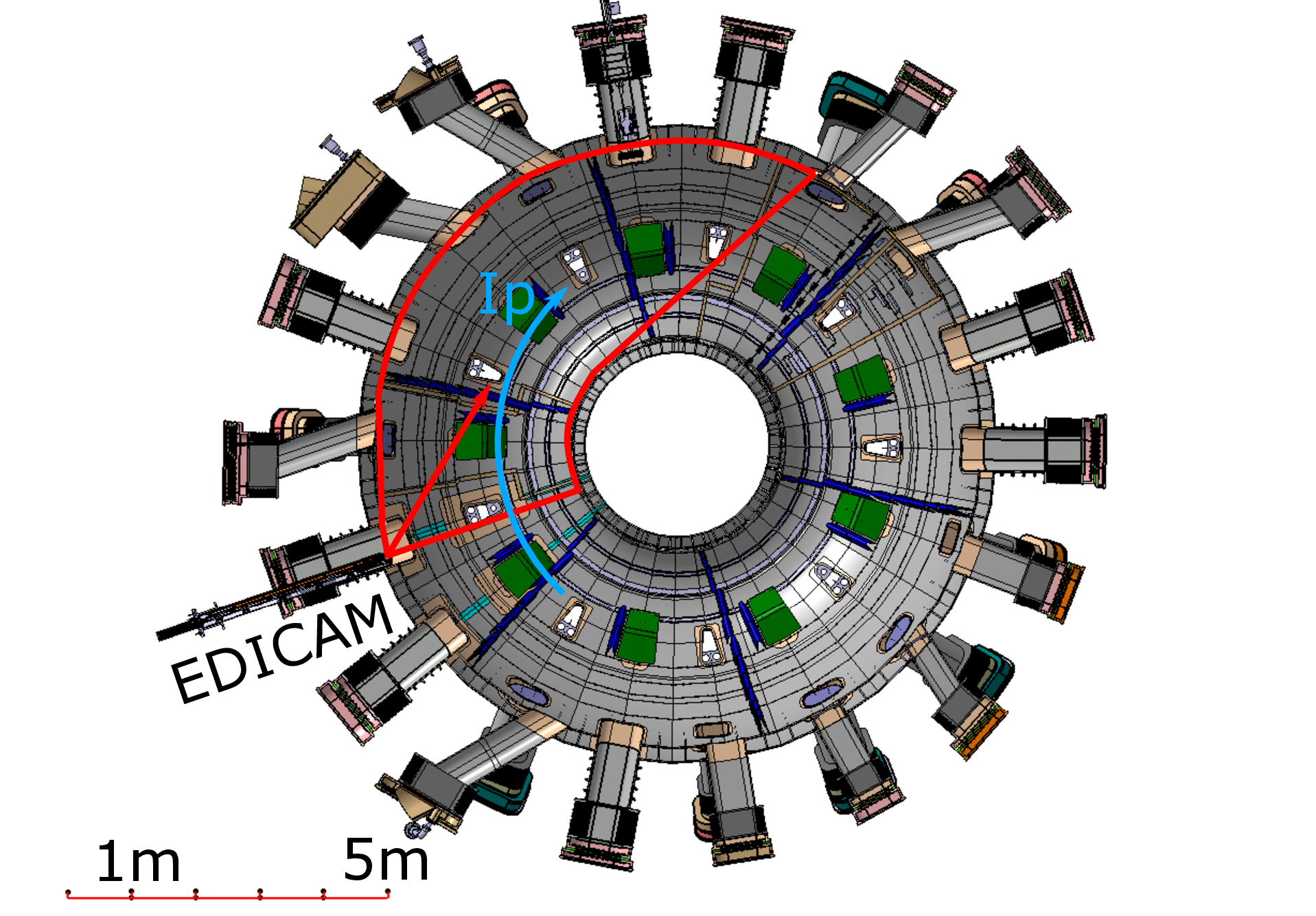}
\caption{Schematic top-view of the EDICAM camera viewing direction, field of view (FOV), and the direction of the toroidal plasma current ($I_p$).}
\label{location}
\end{center}
\end{figure*}

\begin{figure*}
\begin{center}
\includegraphics[scale=0.3]{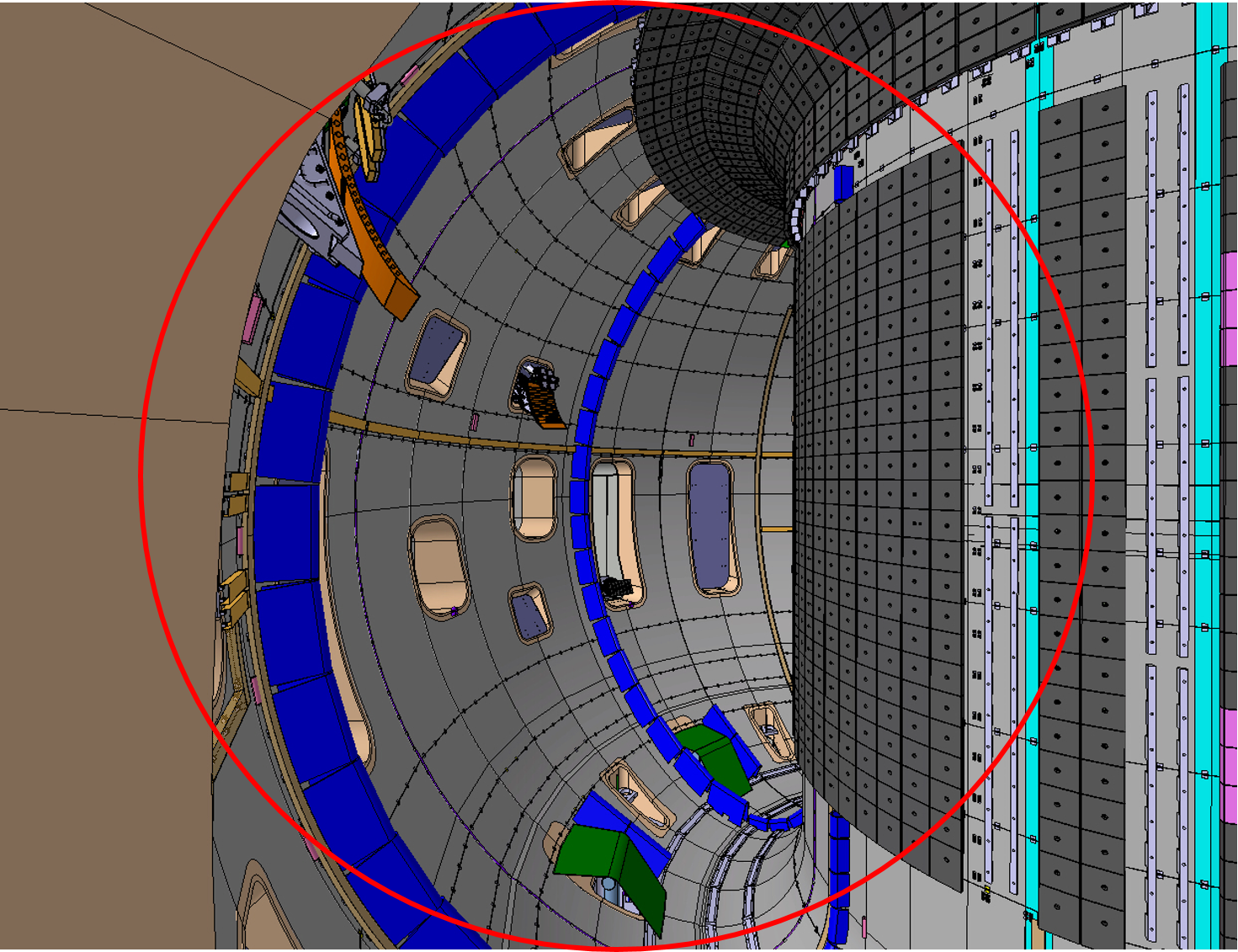}
\caption{The simulated camera view of the vacuum vessel from the EDICAM location. The red circle is the 80-degree field of view seen by the camera system.}
\label{view}
\end{center}
\end{figure*}

The details of the camera characteristics can be found in~\cite{Zoletnik13}, and the JT-60SA-specific installation parameters in~\cite{Szepesi20}. The camera has a $1.3\;\rm Mpixel$ ($1280 \times 1024$) CMOS sensor with $400\; \rm Hz$ maximum readout speed; however, it is capable of non-destructive readout and observation of smaller, specific regions of interest (ROI) with frame rates up to several $10\; \rm kHz$, depending on the resolution~\cite{Szepesi20}. The camera dynamic range is about 10.5 bits, hence the pixels are digitized to 12 bits~\cite{Zoletnik18}. The EDICAM system on JT-60SA does not include a filter, but it is possible to add if deemed necessary. The non-destructive readout allows for the reading of the sensor's content without erasing it. The optical parameters relevant to the modelling are listed in table~\ref{param}. These show that EDICAM represents a typical general purpose visible camera with a wide field of view in a tangential direction into the tokamak from a position close to the mid-plane. The spectral sensitivity of the camera system is shown in Fig.~\ref{spectral}. It can be seen that the EDICAM is most effective in the $520$ to $720\;\rm nm$ spectral range. A sketch of the camera position and viewing direction relative to the torus is shown in Fig.\ref{location} along with the direction of the plasma current. As can be seen, the camera viewing direction and corresponding plasma current direction do not prohibit the detection of the synchrotron radiation emitted by runaway electrons. The simulated camera image from the EDICAM location is shown in Fig.~\ref{view}. The above mid-plane position of the camera necessitates a more detailed study on the actual feasibility of synchrotron detection, which motivates the rest of the paper.

\begin{table}
\centering
\caption{Optical parameters of the EDICAM system as installed on JT-60SA. The coordinate system used for the position has its origin at the symmetry axis of the tokamak at mid-plane.}
\begin{tabular}{l l}
\hline
\textbf{Parameter} & \textbf{Value} \\ \hline
\textbf{Position (x,y,z) ($\rm \bf m$)} & (-4.5304, -1.7552, 0.2312) \\ \hline
\textbf{Viewing direction (vector) ($\rm \bf m$)} & (0.60915, 1.01379, 0) \\ \hline
\textbf{Field of view (FOV) ($\rm \bf degrees $)} & 80 \\ \hline
\textbf{Spectral range ($\rm \bf nm$)} & 520--720 \\ \hline
\textbf{Entrance pupil (diameter) ($\rm \bf mm$)} & 5 \\ \hline
\end{tabular}
\label{param}
\end{table}

\section{Modelling tools}\label{sec:models}
This section describes the modelling tools used in this study: first the DREAM~\cite{Hoppe21} code, used for the disruption simulation, then the SOFT code, used for the modelling of the expected synchrotron radiation image.

To assess the feasibility of the EDICAM system for runaway radiation detection, a JT-60SA-like disruption was simulated with the DREAM code~\cite{Hoppe21}. DREAM is a runaway electron modelling tool developed for the study of disruption runaway electrons. It calculates the evolution of the electron population self-consistently along with the background plasma. The electrons can be handled with different levels of sophistication, ranging from treating all electrons using a fluid model to resolving all electrons kinetically.

The background plasma evolution includes the temperature evolution due to ohmic heating, radiation losses, and collisional energy exchange. The density evolution of the plasma constituent elements is calculated, including both the main ions and the impurities and their charge state evolution. The electric field and the current density evolution are calculated by solving Ampère's and Faraday's laws.~\cite{Hoppe21}.

In the current work, DREAM was used in a kinetic mode, meaning that the entire electron population was evolved by the bounce-averaged Fokker-Planck equation with a relativistic test particle collision operator~\cite{Hoppe21}. This calculates the electron distribution function on a separate 3D phase-space momentum grid for the thermal and suprathermal electron populations and a grid for the runaway electron population. The Dreicer~\cite{Dreicer60} and hot-tail generation of runaway electrons are calculated from the collisional effects, and the Rosenbluth-Putvinski source term~\cite{Rosenbluth97} is used for the avalanche generation mechanism.

The SOFT synthetic diagnostic framework~\cite{Hoppe18_2} was used to calculate the synchrotron radiation expected from the electron distribution function as seen by the EDICAM camera system. The camera parameters listed in table~\ref{param} were added to the model, along with the runaway distribution function calculated by DREAM. The synchrotron radiation was calculated using the cone approximation~\cite{Hoppe18_2}. In this approach, the synchrotron radiation is assumed to be along the guiding centre trajectory of the runaway particles in a hollow cone shape with an infinitely thin side and an opening angle $\theta_p$, the particle pitch angle. In the presented simulations, SOFT modelled the tokamak geometry as a circular torus from the major and minor radii, and a quadratic q-profile was assumed.
\begin{equation}
q(a)=q_1a^2 + q_2,
\end{equation}
where $a$ is the normalized minor radius, $q_1$ and $q_2$ are constants. The q-profile parameters were chosen to resemble the profile calculated by DREAM at the last time point of the simulation, hence $q_1$ was set to $1.23$ and $q_2$ was set to $0.8$. 

\section{Modelling results}\label{sec:Results}

The disruption simulation and the resulting radiation image are presented next. The main plasma parameters, the runaway electron distribution function, and the results of the SOFT simulation are shown, along with a discussion of the outstanding features. The aim of the disruption simulation was to produce a high energy but realistic runaway electron beam in JT-60SA geometry that can be used as an input for the synchrotron image simulation. Accordingly, the disruption simulation input parameters, such as the input argon amount for example, were chosen to produce a significant and energetic runaway electron population, and the scenario is not optimized for disruption mitigation purposes.

\subsection{Disruption simulation}
\begin{figure*}
\begin{center}
\includegraphics[scale=0.25]{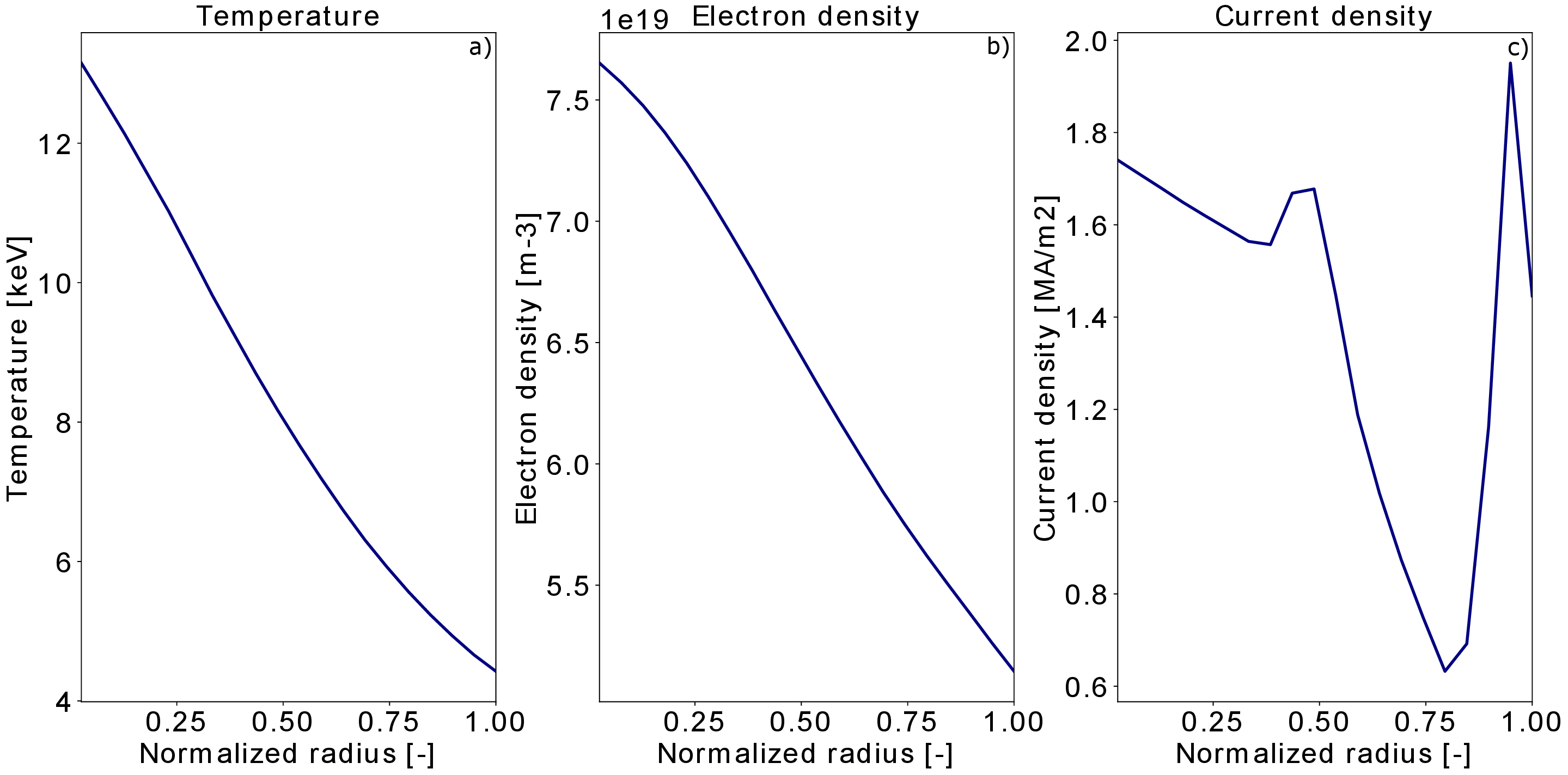}
\caption{The initial plasma profiles. Graph a) shows the temperature profile in keV, graph b) shows the electron density profile, and graph c) shows the current density profile. The sharpness of the peak of the current density at the edge is the result of the interpolation from the EFIT data to the 20 radial point grid of DREAM.}
\label{pre-disruption}
\end{center}
\end{figure*}

The DREAM code was used to simulate a massive material injection (MMI) induced disruption. The initial current density profile was taken from the simulation of Scenario 2 of the JT-60SA research plan~\cite{Giruzzi19}, with a plasma current of $5.5\;\rm MA$. This was assumed to be fully ohmic current. The density and temperature profiles used by the EFIT~\cite{Lao85} magnetic equilibrium calculation were used as initial conditions in the DREAM modelling. The magnetic field strength on axis, major and minor radii were also chosen based on Scenario 2 of the JT-60SA research plan and was set to $2.25\; \rm T$, $2.96\; \rm m$ and $1.18\; \rm m$ respectively. The wall distance was set to 25cm, estimated from values used on JET~\cite{Guillemaut17} and for the wall time 150ms was used~\cite{Takechi19}. The EFIT data was interpolated to the radial grid of DREAM, which was set to 20 radial grid points. The basic pre--disruption plasma parameters are shown in Fig.~\ref{pre-disruption}.

The simulation was done in five separate phases. First, the initial electric field was calculated to give the desired initial plasma current based on the given density and temperature profiles. Then $10^{20}\; \rm m^{-3}$ argon gas density was introduced to the plasma uniformly at every radial point. The introduced Ar density is comparable to the plasma density. The transient ionization dynamics were resolved using a very short time step, lasting $1\; \rm \mu s$ after which the prescribed exponential temperature decay started. This temperature decay was further simulated on a longer timescale to reach below $100\; \rm eV$ temperatures. It was artificially prescribed at every radial point until the temperature reached $100\; \rm eV$ at the innermost radial point. After the prescribed phase ended, the temperature evolution was calculated self-consistently, mainly governed by the radiation of the injected argon gas and the ohmic heating of the plasma current. The simulation finished with the calculation of the runaway plateau phase, during which the electron population can experience significant pitch angle scattering~\cite{Hoppe21_2}. The total simulated time was about $8.6\;\rm ms$.  

\begin{figure*}
\begin{center}
\includegraphics[scale=0.25]{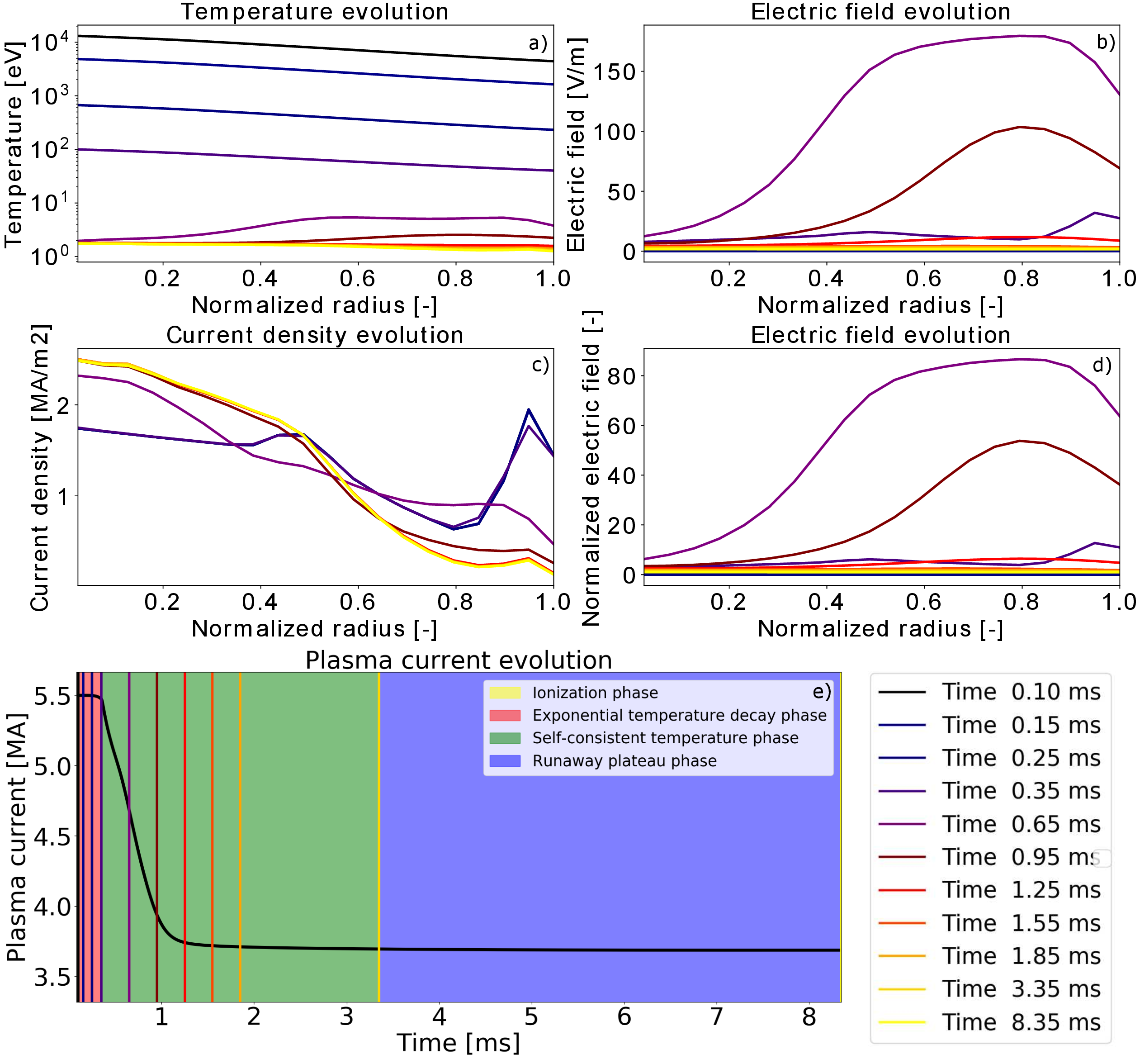}
\caption{The time evolution of the main plasma parameters. Plot a) shows the temperature, and plot b) shows the electric field evolution. The change in the current density is plotted in graph c), while the electric field normalized to the effective critical electric field is plotted in graph d). Graph e) shows the evolution of the plasma current as a function of time. The different simulation phases are indicated with shaded regions, while the times plotted on the first four graphs are shown with vertical lines of the corresponding color. Note the logarithmic vertical axis on the temperature plot compared to the other plots. The color going from dark to light shows the passing of time.}
\label{params}
\end{center}
\end{figure*}

The temperature as a function of radius at different times throughout the simulation is shown in Fig.~\ref{params}a. The initial temperature at the magnetic axis is $13.4\; \rm keV$. It can be seen that until the temperature reaches $100\;\rm eV$ at the centre, the temperature drops exponentially at every radial point, keeping the initial profile shape. Below $100\;\rm eV$ the temperature is calculated using an energy-balance model, and it equilibrates along the radius. The plasma cools slower at the outer regions, and simultaneously the current density, shown in Fig.~\ref{params}c, relaxes at a slower pace, as seen around $0.7$ normalized radius. The final current density is completely provided by runaway electrons, generated by the electric field induced during the disruption. The accelerating electric field can be seen in Fig.\ref{params}b, while the electric field normalized to the effective critical field~\cite{Hesslow18} is shown in Fig.~\ref{params}d. The overall shape of the two electric field plots is similar since the effective critical field is mainly dependent on the density and weakly on the temperature. As the density profile is mostly flat during the simulation, the critical field will be similar at every radial point, and the normalization will preserve the shape of the electric field. In Fig.~\ref{params} a very high electric field can be seen around $0.65\;\rm ms$ when the temperature dropped to below $10\;\rm eV$. This peak is off-axis, located at about $0.75$ normalized radius due to the decay of the current peak seen in Fig.~\ref{params}b. This electric field accelerates the electrons further and yields a more energetic runaway population at this location.

The plasma current evolution is plotted in Fig.~\ref{params}, with the different simulation phases indicated with shaded regions. The ionization of the injected argon was simulated in $1\;\rm \mu s$, so the yellow shaded area is difficult to see behind the vertical lines showing the plot times from the main plasma profiles shown in Fig.\ref{params} a)-d). The thermal quench is simulated between the times shaded red with exponentially decaying plasma temperature. The current quench is finished by $1\; \rm ms$ with a current of less than $4\; \rm MA$. In the runaway plateau phase, the plasma current decays insignificantly on the simulated time scale. The runaway electrons are not significantly accelerated to higher energies during this phase, as the accelerating electric field is proportional to the current decay time. The current drops much faster in the current quench phase, hence the acceleration in the runaway plateau phase should be negligible. A sufficient runaway electron population was achieved at the end of the simulation, so the plasma evolution was not calculated further.

\begin{figure*}
\begin{center}
\includegraphics[scale=0.25]{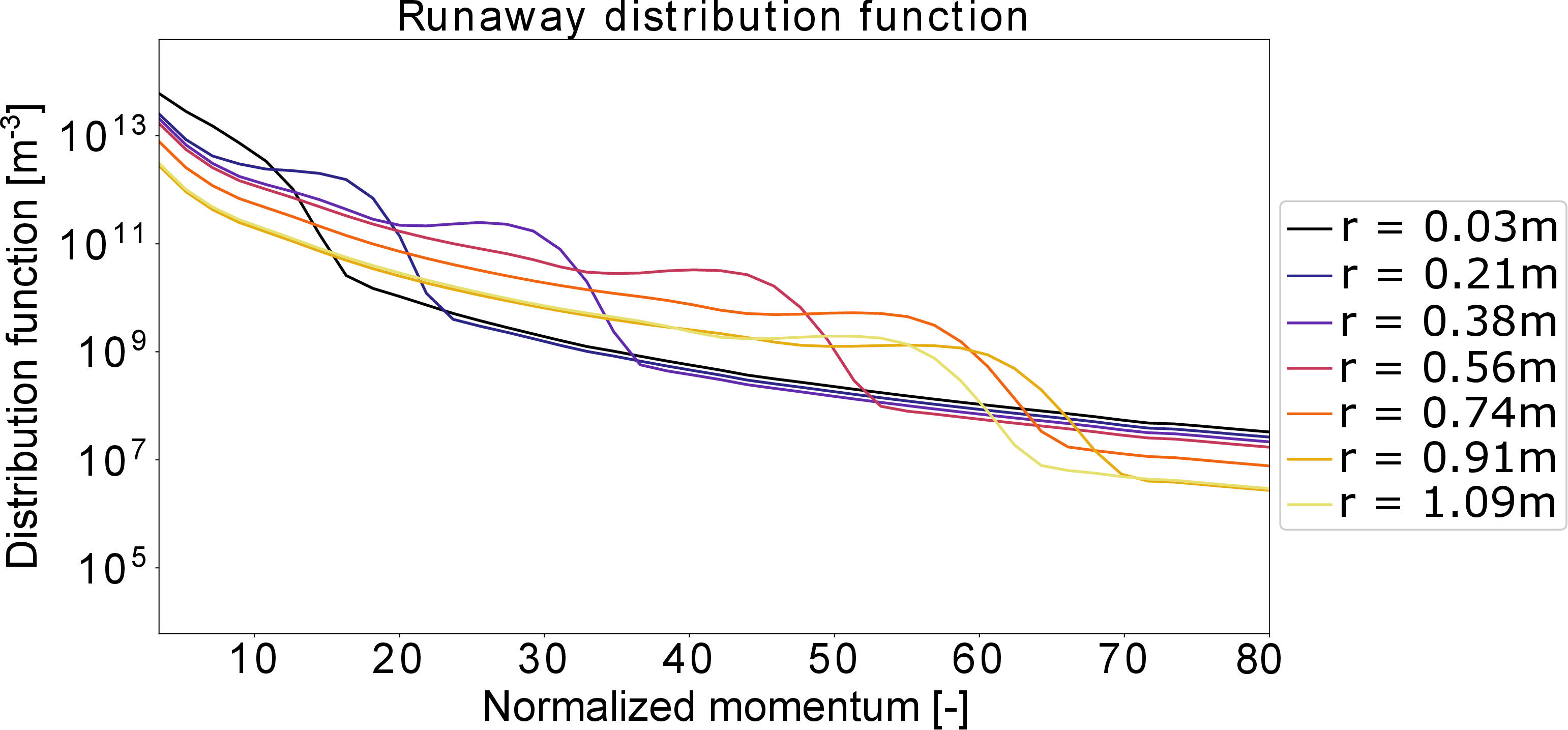}
\caption{The runaway electron distribution function plotted for several radial grid points at the last time step as a function of momentum $p$ normalized to $m_ec$. The highest energy runaway electron population is located at the edge of the plasma. The color going from dark to light indicates the radial position, going from the centre to the edge.}
\label{dist}
\end{center}
\end{figure*}

The angle-averaged distribution function at the end of the simulation is shown in Fig.~\ref{dist} for several simulated radial locations. The hot-tail seed formed during the thermal quench can be seen accelerated to higher energies, as seen by the local peaks of the distribution function at various points at different radial coordinates. The maximum runaway electron momentum reached is about $65m_ec$ or about $33\; \rm MeV$ at about $0.91\; \rm m$ away from the magnetic axis, corresponding to about $0.77$ normalized radius. The runaway distribution in the centre only reached about 10-$20m_ec$, as the electric field did not penetrate the plasma deep enough to accelerate the population at the magnetic axis, as seen in Fig.\ref{params} b) and d). The synchrotron radiation is highly dependent on the runaway energy and pitch angle, so the radiation is expected to be mainly off-axis.

\subsection{Simulation of the radiation image}
\begin{figure*}
\begin{center}
\includegraphics[scale=0.5]{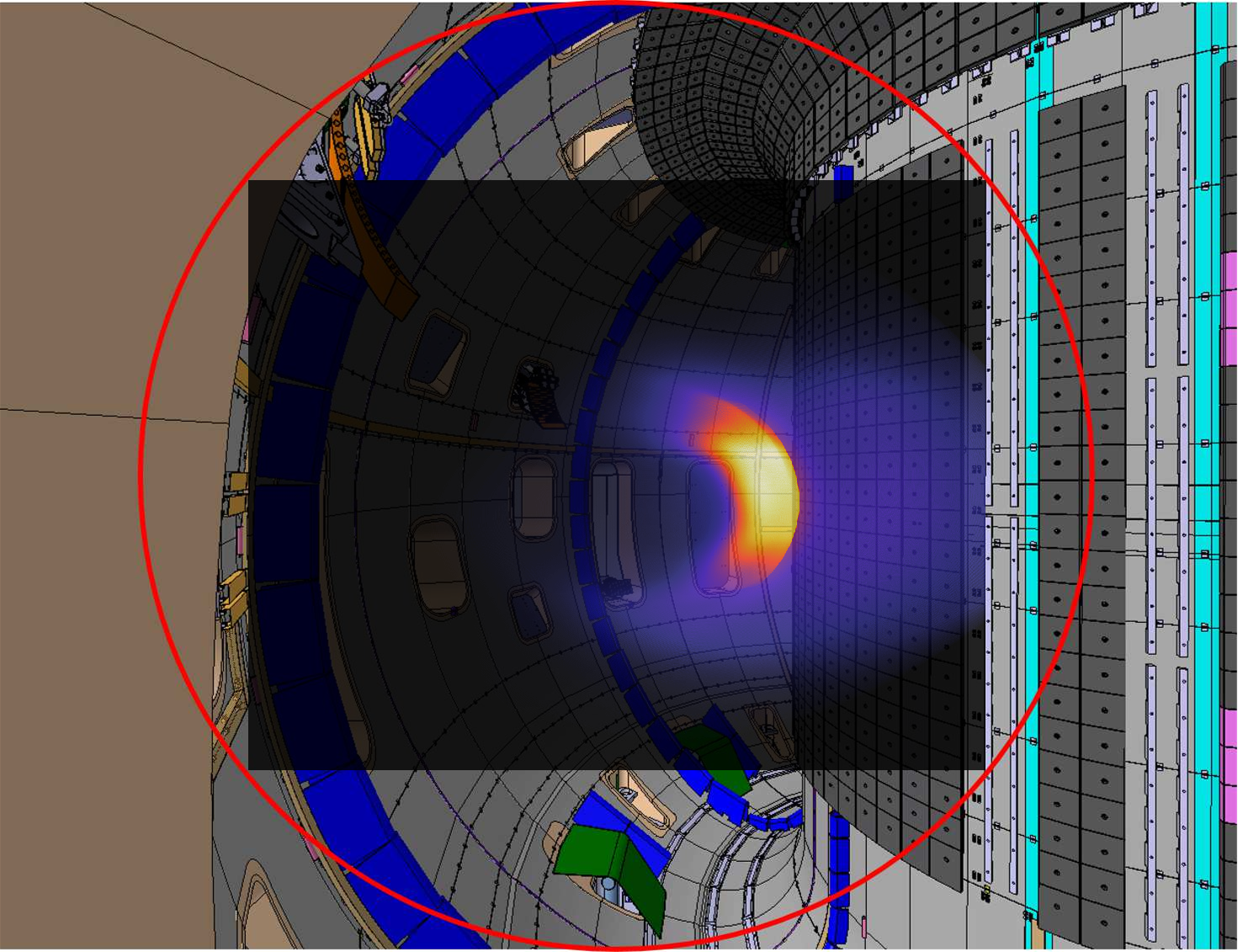}
\caption{The synchrotron radiation of the runaway beam during a JT-60SA-like disruption as seen by the EDICAM visible camera system, plotted to the camera picture shown in Fig.~\ref{view}.}
\label{radiation}
\end{center}
\end{figure*}

The runaway electron distribution was given to the SOFT code, which calculated the synchrotron radiation as seen by the EDICAM camera. The simulated camera image is shown in Fig.~\ref{radiation}. Significant radiation can be seen on the high-field side in a crescent-like shape, while the radiation is negligible at the magnetic axis. The simulated image resembles observations of synchrotron radiation emitted by runaway electrons in other tokamaks~\cite{Reux21}.

The effect of vertical displacement of the runaway electron beam was investigated by moving the camera location in the SOFT simulations. It was found that the displacement of the beam did not affect the detection capability of the camera system, but changed the location of the radiation spot.

\begin{figure*}
\begin{center}
\includegraphics[scale=0.28]{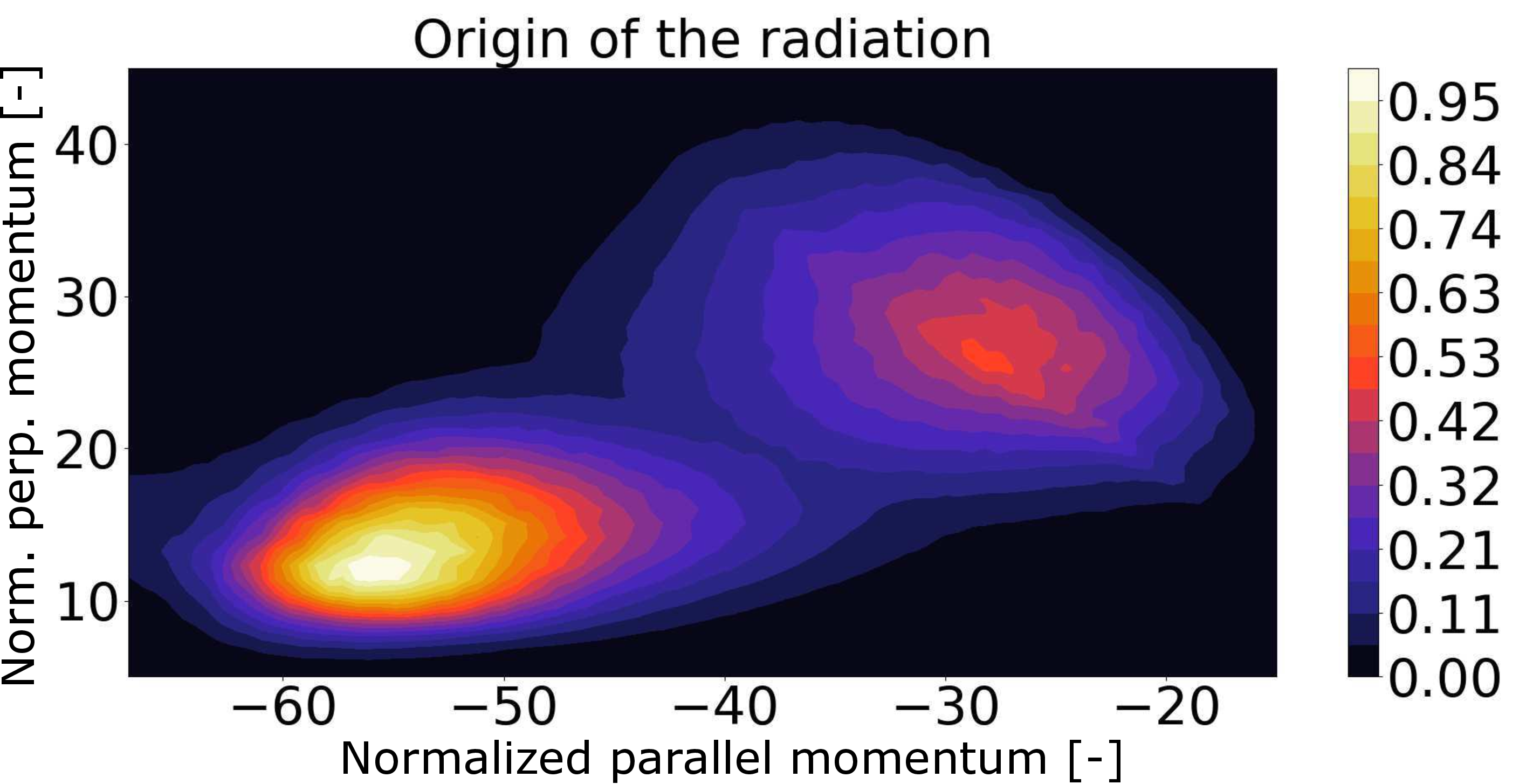}
\caption{The Green's function from the SOFT simulation weighted with the distribution function from DREAM. The peak of synchrotron radiation origin is at $p_\parallel \approx 57m_ec$ with $p_{\perp} \approx 12m_ec$. The colorbar is normalized to the maximum of the peak.}
\label{green}
\end{center}
\end{figure*}

The origin of the radiation in momentum space is illustrated by Fig.\ref{green}, where the Green's function weighted with the DREAM distribution function is plotted as a function of parallel and perpendicular momenta normalized to $\rm m_ec$. The Green's function contains information on the tokamak geometry and the momentum dependence of the radiation. It can be seen that most of the radiation originates at $p=57m_ec$. This location corresponds to the highest energies of the runaway electron distribution function seen in Fig.\ref{dist}. The radiation is also sensitive to the pitch angle of the particles, which explains the shape of the maximum in Fig.~\ref{green}. There is a ridge extending to $30\; m_ec$ which is most likely a result of the runaway electron population at the inner radii. The dip between the two maxima is probably an artefact of the interpolation between the DREAM and the SOFT grids. The dominant radiation spot corresponds to the high-energy runaway population at about $0.75$ normalized radius. The hollow radiation shape can be explained by the enhancing effect of the high magnetic field strength on the synchrotron radiation as well as the high runaway electron energies at larger radii. Since more and more energetic runaway electrons are generated at the edge, the main source of radiation is coming from the edge on the high-field side of the tokamak.

\section{Conclusions}\label{sec:Conclusions}
The DREAM and SOFT codes were used to simulate the synchrotron radiation originating from a runaway population generated in a JT-60SA-like disruption as seen by the EDICAM visible camera system. DREAM calculated the runaway electron population during an argon-induced disruption. The resulting runaway distribution was given to SOFT to calculate the synchrotron radiation to assess the feasibility of the EDICAM camera system for synchrotron radiation detection. It was found that the synchrotron radiation from runaway electrons is emitted in the right direction to be observed by the installed EDICAM system. Moreover, the EDICAM characteristics make it ideal to observe fast phenomena in tokamaks, such as disruptions. Our main conclusion is that the EDICAM camera system is capable of detecting synchrotron radiation from high-energy runaway electrons generated in major disruptions. 

In this feasibility study, the argon injection simulated by a simple model of uniform argon density introduced simultaneously to the plasma at every radial point. The simulation could be improved by considering a more realistic injection of the argon gas, but it would most likely change the radiation shape, not the radiation detection capability of the camera system. The radiation image could also be improved by adding a realistic camera view, such as the exact shape of the tokamak, the effect of reflections, background radiation to the SOFT code or another model, but it is beyond the scope of this study. The present results might motivate a more detailed study of the range of applicability of the EDICAM system, in which the above effects could be investigated, and its applicability for early detection of runaway electron beams in mitigated disruptions and other operation regimes.

    \section*{Acknowledgments} 
The authors are grateful to N. Hajnal for the simulated camera image of the EDICAM camera system, and the camera location image and O. Asztalos for the EFIT data provided for the input to DREAM. This work has been carried out within the framework of the EUROfusion Consortium, funded by the European Union via the Euratom Research and Training Programme (Grant Agreement No 101052200 — EUROfusion). Views and opinions expressed are however those of the author(s) only and do not necessarily reflect those of the European Union or the European Commission. Neither the European Union nor the European Commission can be held responsible for them. G.~I.~Pokol and S.~Olasz acknowledge the support of the National Research, Development and Innovation Office (NKFIH) Grant FK132134. This work was supported in part by the Swiss National Science Foundation.





\end{document}